\documentclass[11pt,a4paper]{article}
\pdfoutput=1

\usepackage[utf8]{inputenc}
\usepackage[english]{babel}
\usepackage{chicago}

\usepackage[round]{natbib}
\usepackage[margin=1in]{geometry}

\usepackage{graphicx, subfig, caption, a4wide, booktabs, fancyvrb, upquote, longtable, multirow, enumitem, ifsym, hyperref}
\usepackage{amsmath, amsfonts, amssymb}

\usepackage{afterpage}

\usepackage{color}
\definecolor{darkgrey}{rgb}{0.2,0.2,0.2}
\usepackage{fancyvrb}
\DefineVerbatimEnvironment{Code}{Verbatim}
  {fontsize=\normalsize,
   formatcom=\color{darkgrey},
   baselinestretch=1}
   
\newcommand{\x}{\boldsymbol{x}}

\newcommand{\thetab}{\boldsymbol{\theta}}
\newcommand{\PSI}{\boldsymbol{\Psi}}
\newcommand{\mub}{\boldsymbol{\mu}}
\newcommand{\Sigmab}{\boldsymbol{\Sigma}}

\newcommand{\W}{\boldsymbol{W}}
\newcommand{\Wcrit}{\mathcal{W}}
\newcommand{\X}{\boldsymbol{X}}
\newcommand{\U}{\boldsymbol{U}}
\newcommand{\D}{\boldsymbol{D}}
\renewcommand{\O}{\boldsymbol{O}}
\newcommand{\A}{\boldsymbol{A}}
\newcommand{\V}{\boldsymbol{V}}
\renewcommand{\u}{\boldsymbol{u}}
\renewcommand{\v}{\boldsymbol{v}}
\newcommand{\Z}{\boldsymbol{Z}}
\newcommand{\I}{\boldsymbol{I}}
\newcommand{\0}{\boldsymbol{0}}
\newcommand{\1}{\boldsymbol{1}}
\renewcommand{\S}{\boldsymbol{S}}
\newcommand{\xbar}{\bar{\boldsymbol{x}}}
\newcommand{\T}{{}^{\top}}
\renewcommand{\hat}[1]{\widehat{#1}}
\newcommand{\XX}{\mathbb{X}}
\DeclareMathOperator{\Normal}{\mathrm{N}}
\DeclareMathOperator{\diag}{\mathrm{diag}}
\DeclareMathOperator{\Exp}{\mathrm{E}}
\DeclareMathOperator{\Var}{\mathrm{Var}}
\DeclareMathOperator{\Cor}{\mathrm{Cor}}
\newcommand{\Zstd}{\Z_{\text{STD}}}
\newcommand{\Zsph}{\Z_{\text{SPH}}}
\newcommand{\Zpcs}{\Z_{\text{PCS}}}
\newcommand{\Zpcr}{\Z_{\text{PCR}}}
\newcommand{\Zsvd}{\Z_{\text{SVD}}}
\newcommand{\Model}{\mathcal{M}}
\newcommand{\BIC}{\mathrm{BIC}}

\makeatletter
\newcommand\code{\bgroup\@makeother\_\@makeother\~\@makeother\$\@codex}
\def\@codex#1{{\normalfont\ttfamily\hyphenchar\font=-1 #1}\egroup}
\makeatother
\newcommand{\proglang}[1]{\textsf{#1}}
\newcommand{\pkg}[1]{\code{#1}}

\usepackage{fancyvrb}
\DefineVerbatimEnvironment{Code}{Verbatim}
  {fontsize=\footnotesize,
   formatcom=\color{darkgrey},
   baselinestretch=0.9}
   
\begin{document}


\title{Improved initialisation of model-based clustering
       using Gaussian hierarchical partitions}
\author{%
Luca Scrucca \\ Universit\`a degli Studi di Perugia \and
Adrian E. Raftery \\ University of Washington
\thanks{Luca Scrucca is Assistant Professor of Statistics, Dipartimento di Economia, Universit\`a degli Studi di Perugia, Via A. Pascoli, 20, 06123 Perugia, Italy; Email: luca@stat.unipg.it.
Adrian E. Raftery is Professor of Statistics and Sociology, Department of Statistics, University of Washington, Box 354322, Seattle, WA 98195-4322, USA; Email: raftery@u.washington.edu.
Raftery's research was supported by the Eunice Kennedy Shriver National Institute of Child Health and Development through grants nos. R01 HD054511 and R01 HD070936, and by a Science Foundation Ireland E.T.S.~Walton visitor award, grant reference 11/W.1/I2079.}}
\date{\today}
\maketitle

\begin{abstract}
Initialisation of the EM algorithm in model-based clustering is often crucial.
Various starting points in the parameter space often lead to different local maxima of the likelihood function and, so to different clustering partitions. 
Among the several approaches available in the literature, model-based agglomerative hierarchical clustering is used to provide initial partitions in the popular \pkg{mclust} \proglang{R} package. 
This choice is computationally convenient and often yields good clustering partitions. However, in certain circumstances, poor initial partitions may cause the EM algorithm to converge to a local maximum of the likelihood function. We propose several simple and fast refinements based on data transformations and illustrate them through data examples. \\

\end{abstract}

\newpage
\baselineskip=18pt


\section{Introduction}

Model-based clustering is an increasing popular method for unsupervised learning. 
In contrast to classical heuristic methods, such as $k$-means and hierarchical clustering, model-based clustering methods rely on a probabilistic assumption about the data distribution. 
According to the main underlying assumption, data are generated from a mixture distribution, where each cluster is described by one or more mixture components. 
Maximum likelihood estimation of parameters is usually carried out via the EM algorithm \citep{Dempster:Laird:Rubin:1977}.
The EM algorithm is an iterative, strictly hill-climbing procedure whose performance can be very sensitive to the starting point because the likelihood surface tends to have multiple modes, although it usually produces sensible results when started from reasonable starting values. 
Thus, good initialisation is crucial for finding MLEs, although no method suggested in the literature uniformly outperforms the others. 

In the case of Gaussian model-based clustering \citep{Fraley:Raftery:2002}, several approaches are available, both stochastic and deterministic, for selecting an initial partition of the observations, or an initial estimate of the parameters. 
In the \pkg{mclust} \proglang{R} package \citep{Fraley:Raftery:Murphy:Scrucca:2012, Rpkg:mclust}, the EM algorithm is initialised using the partitions obtained from model-based agglomerative hierarchical clustering. Efficient numerical algorithms exist for approximately maximise the classification likelihood with multivariate normal models. However, in certain circumstances, poor initial partitions may cause the EM algorithm to converge to a local maximum of the likelihood function.

In this contribution we discuss cases where an initial partition may lead to sub-optimal maximum likelihood estimates when applied to coarse data with ties (e.g. discrete data or continuous data that are rounded in some form when measured), and we present some possible refinements to improve the fitting of such finite mixture models.

The outline of this article is as follows. 
Section~\ref{sec:background} gives a brief review of background material on model-based clustering, with special attention devoted to some of the proposals available in the literature for the initialisation of EM algorithm.
Section~\ref{sec:mbhac} discusses the model-based hierarchical agglomerative clustering method used for starting the EM algorithm. This is a very convenient and efficient algorithm, but in certain circumstances presents a serious drawback. 
Section~\ref{sec:mbhact} contains some simple transformation-based methods to refine the EM initialisation step derived from model-based agglomerative hierarchical clustering. 
The behaviour of these methods is illustrated through the use of real data examples in Section~\ref{sec:analyses}. 
The final section provides some concluding remarks.

\section{Background}
\label{sec:background}

\subsection{Model-based clustering overview}

Let $\x_1,\x_2,\ldots,\x_n$ be a sample of $n$ independent identically distributed observations. 
The distribution of every observation is specified by a probability mass or density function through a finite mixture model of $G$ components, which takes the following form
\begin{equation}
f(\x; \PSI) = \sum_{k=1}^G \pi_k f_k(\x; \thetab_k),
\label{eq:finmixdens}
\end{equation}
where $\PSI = \{\pi_1, \ldots, \pi_{G-1}, \thetab_1, \ldots, \thetab_G\}$ are the parameters of the mixture model, $f_k(\x; \thetab_k)$ is the $k$th component density at $\x$ with parameter(s) $\thetab_k$, $(\pi_1,\ldots,\pi_{G-1})$ are the mixing weights or probabilities (such that $\pi_k > 0$, $\sum_{k=1}^G\pi_k = 1$), and $G$ is the number of mixture components.

Assuming $G$ fixed, mixture model parameters $\PSI$ are usually unknown and must be estimated. 
The log-likelihood function corresponding to equation \eqref{eq:finmixdens} is given by $\ell(\PSI; \x_1, \ldots, \x_n) = \sum_{i=1}^{n} \log(f(\x_i; \PSI))$.
Direct maximisation of the log-likelihood function is often complicated, so  MLE of finite mixture models is usually carried out via the EM algorithm \citep{McLachlan:Peel:2000}.

In the model-based approach to clustering, each component of a finite mixture of density functions belonging to a given parametric class is associated with a group or cluster. 
Most applications assume that all component densities arise from the same parametric distribution family, although this need not be the case in general.
A popular model assumes a Gaussian distribution for each component, i.e. $f_k(\x; \thetab_k) \sim \Normal(\mub_k, \Sigmab_k)$.
Thus, clusters are ellipsoidal, centred at the mean vector $\mub_k$, and with other geometric features, such as volume, shape and orientation, determined by $\Sigmab_k$.
Parsimonious parameterisations of the covariance matrices can be defined by means of eigen-decomposition in the form $\Sigmab_k = \gamma_k \O_k \A_k \O\T_k$, where $\gamma_k$ is a scalar controlling the volume of the ellipsoid, $\A_k$ is a diagonal matrix specifying the shape of the density contours, and $\O_k$ is an orthogonal matrix which determines the orientation of the corresponding ellipsoid \citep{Banfield:Raftery:1993, Celeux:Govaert:1995}. \citet[Table~1]{Fraley:Raftery:Murphy:Scrucca:2012} summarized some parameterisations of within-group covariance matrices available in the \pkg{mclust} software, and the corresponding geometric characteristics.

The number of mixture components and the parameterisation of the component covariance matrices can be selected on the basis of model selection criteria, such as the Bayesian information criterion \citep[BIC;][]{Schwartz:1978, Fraley:Raftery:1998} or the integrated complete-data likelihood criterion \citep[ICL;][]{Biernacki:Celeux:Govaert:2000}.

\subsection{Initialisation of EM algorithm}

The EM algorithm is an easy to implement, numerically stable algorithm, which has, under fairly general conditions, reliable global convergence. However, it may converge slowly and, like any other Newton-type method, does not guarantee convergence to the global maximum when there are multiple maxima \citep[p. 29]{McLachlan:Krishnan:2008}. Further, in the case of finite mixture modelling, the estimates obtained depends on the starting values. 
Thus initialisation of EM is crucial because the likelihood surface tends to have multiple modes, although it usually produces sensible results when started from reasonable starting values \citep[p. 150]{Wu:1983, Everitt:2011}.

Several approaches are available, both stochastic and deterministic, for initialising the EM algorithm. 
Broadly speaking, there are two general approaches. The first one starts from some initial values for the parameters to be estimated. A simple strategy is based on generating several candidates by drawing parameter values uniformly at random over the feasible parameters regions. Since the random-starts strategy has a fair chance of not providing good initial starting values, a common suggestion to alleviate this problem is to run the EM algorithm with several random starts and to choose the best solution. However, such a strategy can be quite time consuming and is not always practical, especially for high-dimensional datasets. 

Two other stochastic initialisation schemes are the so-called \textit{em}EM and \textit{rndEM}. The former approach, proposed by \citet{Biernacki:etal:2003}, uses several short runs of the EM initialised with valid random starts as parameter estimates until an overall number of total iterations is exhausted. Then, the solution with the highest log-likelihood is chosen to be the initialiser for the long EM, which runs until the usual strict convergence criteria are met. This approach is computationally intensive and the same comments made above about
random starts apply to it also.

Two related approaches were proposed by \citet{Maitra:2009}, one called \textit{Rnd-EM} where the short EM stage is replaced by choosing multiple starting points and evaluating the log-likelihood at these values without running any EM iterations. Then, the best obtained solution serves as an initialiser for the long EM stage. The second proposal is a staged approach based on finding a large number of local modes of the dataset, and then to choose representatives from the most widely-separated ones.
This approach is reported to be very time-consuming for high-dimensional data, and \citet{Melnykov:Maitra:2010} found the method to be outperformed by \textit{em}EM and \textit{RndEM}.
Recently, \citet{Melnykov:Melnykov:2012} have proposed a strategy for initialising mean vectors by choosing points with higher concentrations of neighbours and using a truncated normal distribution for the preliminary estimation of dispersion matrices. 

A second kind of approach for initialising the EM algorithm is based on the partition obtained from another clustering algorithm, e.g. $k$-means or  hierarchical agglomerative clustering (HAC). 
In this case, the final classification is used to start the EM algorithm from the M-step. Unfortunately, most of these partitioning algorithm have several drawbacks, such as the need to be properly initialised or the tendency to impose specific shapes and patterns on clusters.
In the popular \pkg{mclust} package for \proglang{R}, the EM algorithm is initialised using the partitions obtained from model-based hierarchical agglomerative clustering (MBHAC). In this approach, $k$ clusters are obtained from a large number of smaller clusters by recursively merging the two clusters that have the smallest dissimilarity in a model-based sense, i.e. the dissimilarity used for agglomeration is derived from a probabilistic model. \citet{Banfield:Raftery:1993} proposed a dissimilarity based on a Gaussian mixture model, which is equal to the decrease in likelihood resulting by the merging of two clusters. \citet{Fraley:1998} showed how the structure of some specific Gaussian models can be exploited to yield efficient algorithms for agglomerative hierarchical clustering.
More details about this approach are discussed in the following section.

\section{Model-based hierarchical agglomerative clustering}
\label{sec:mbhac}


The objective of MBHAC is to obtain a hierarchical ordering of clusters of $n$ objects on the basis of the some measure of similarity among them. The result is a tree-like structure, which proceeds from $n$ clusters containing one object to one cluster containing all $n$ objects by successively merging objects and clusters.

Given $n$ objects or observations $(\x_1, \ldots, \x_n)$, let $(z_1, \ldots, z_n)\T$ denote the classification labels, i.e. $z_i=k$ if $\x_i$ is assigned to cluster $k$. The unknown parameters are obtained by maximising the classification likelihood
$$
\mathcal{L}_C (\thetab_1, \ldots, \thetab_G, z_1, \ldots, z_n|\x_1, \ldots, \x_n) = \prod_{i=1}^n f_{z_i}(\x_i|\thetab_{z_i}) .
$$
Assuming the multivariate Gaussian distribution for $f_{k}(\x_i|\thetab_{k})$, the parameters are the mean vector $\mub_k$ and the covariance matrix $\Sigmab_k$. By imposing different covariance structures over $\Sigmab_k$, different criterion can be derived \citep[see][Table~1]{Fraley:1998}. 
For instance, when the covariance is allowed to be different among clusters, the criterion to be minimised at each hierarchical stage is
\begin{equation}
\Wcrit_G = \sum_{k=1}^G n_k \log \left|\frac{\W_k}{n_k}\right|,
\label{eq:MBHAC}
\end{equation}
where $n_k$ is the number of observations in group $k$ and $\W_k$ is the sample cross-product matrix for the $k$th group ($k=1,\ldots,G$).
Efficient numerical algorithms are available 
for approximately maximising the classification likelihood (or, equivalently, minimising the corresponding criterion) with multivariate Gaussian models have been discussed.

As mentioned, the above MBHAC approach is used for starting the EM algorithm in the \pkg{mclust} \proglang{R} package. This is particularly convenient because the underlying probabilistic model can be shared by the initialisation step and the model fitting step. 
MBHAC is also computationally convenient because a single run provides the basis for initialising the EM algorithm for any number of mixture components and parameterisations of the component covariance matrices.

However, a serious problem for the MBHAC approach may arise when, at any stage, two pairs of objects attain the same minimum value for the criterion in \eqref{eq:MBHAC}. In the presence of coarse data, resulting from the discrete nature of the data or from continuous data that are rounded in some way when measured, ties must be broken by choosing the pair of entities that will be merged. This is often done at random but, regardless of which method is adopted for breaking ties, this choice can have important consequences because it changes the clustering of the remaining observations.
In this case the final EM solution may depend on the ordering of the variables, and to a lesser extent on permutation of the observations (the latter case is not studied further in this paper).

This difficulty is known as the \textit{ties in proximity problem} in the  hierarchical clustering literature \citep[see, for example,][Sec. 3.2.6]{Jain:Dubes:1988}. This problem can also arise in other contexts, such as $k$-means clustering \citep[p. 42]{Gordon:1999} or partition around medoids \citep[PAM;][p. 104]{Kaufman:Rousseeuw:1990}.

\section{Transformation-based approaches for obtaining starting partitions in model-based hierarchical agglomerative clustering}
\label{sec:mbhact}

In this section we describe some simple proposals for starting the EM algorithm using the partitions obtained with MBHAC. Ideally, we would like to retain the positive aspects of such approach, but, at the same time, reduce the chance that a poor initial partition causes the EM algorithm to converge to a local maximum of the likelihood function. 

The idea is to project the data through a suitable transformation before applying the MBHAC at the initialisation step. Once a  reasonable hierarchical partition is obtained, the EM algorithm is run using the data on the original scale. 

Let $\X$ be the $(n \times p)$ data matrix, and $\XX = (\X - \1_n\xbar\T)$ be the corresponding centred matrix, where $\xbar=(\bar{x}_1,\ldots,\bar{x}_p)\T$ is the vector of sample means, and $\1_n$ is the unit vector of length $n$. Let $\hat{\Sigmab} = \{s_{ij}\} = \XX\T\XX/n$ be the $(p \times p)$ sample covariance matrix.
Consider the singular value decomposition (SVD) of the centred data matrix, 
$$
\XX = \U \D \V\T = \sum_{i=1}^r \lambda_i\u_i\v\T_i,
$$ 
where $\u_i$ are the right singular vectors, $\v_i$ the left singular vectors, $\lambda_1 \ge \lambda_2 \ge \ldots \ge\lambda_r > 0$ the singular values, and $r$ the rank of matrix $\XX$. 
Similarly, the centred and scaled data matrix can be decomposed as
$$
\XX\S^{-1/2} = \U^* \D^* \V^*\T = \sum_{i=1}^r \lambda^*_i\u^*_i\v^*_i\T,
$$
where $\S = \diag(s^2_1, \ldots, s^2_p)$ is the diagonal matrix of sample variances.
In both SVD decompositions the rank $r$ is the rank of the data matrix, i.e. $r \le \min(n,p)$, with equality when there are no singularities.

We now provide the details of the transformations investigated and some remarks.

\subsection{Data standardisation}

The first data transformation which has been investigated is the simple standardisation of variables (\code{STD}): 
$$
\Zstd \;\leftarrow\; \XX\S^{-1/2},
$$
for which $\Exp(\Zstd) = \0$ and $\Var(\Zstd) = \Cor(\Zstd)$, i.e. the features are centred at zero, with unit variances and  correlated. This transformation only shifts the data at the origin and scale the variables to a common unit of measure, while preserving the data distribution shape.

\subsection{Data sphering}

Sphering (or whitening) the data (\code{SPH}) is obtained by applying the following transformation:
$$
\Zsph \;\leftarrow\; \XX \V \D^{-1}\sqrt{n} = \U\sqrt{n},
$$
where $\V$ and $\D^{-1}\sqrt{n} = \diag(\sqrt{n}/\lambda_i)$ are, respectively, the matrix of eigenvectors and the diagonal matrix of square root inverse of eigenvalues from the spectral decomposition of the sample marginal covariance, $\hat{\Sigmab}$. For this transformation, $\Exp(\Zsph) = \0$ and $\Var(\Zsph) = \I$, so the features are centred at zero, with unit variances and uncorrelated. Thus, this transformation converts an elliptically shaped symmetric cloud of points into a spherically shaped cloud.

\subsection{PCA scores from covariance matrix}

The principal component transformation from the covariance matrix (\code{PCS}) is obtained as:
$$
\Zpcs \;\leftarrow\; \XX \V = \U \D,
$$
for which $\Exp(\Zpcs) = \0$ and $\Var(\Zpcs) = \D^2/n = \diag(\lambda_i^2/n)$, so the features are centred, uncorrelated and with decreasing variances equal to the eigenvalues of $\hat{\Sigmab}$. Usually the first few components account for most of the dispersion in the data.
A similar idea was proposed by \citet{McLachlan:1988}, who discussed the use of principal component analysis in a preliminary exploratory analysis for the selection of suitable starting values. 

\subsection{PCA scores from correlation matrix}

The principal component transformation from the correlation matrix (\code{PCR}) is defined as:
$$
\Zpcr \;\leftarrow\; \XX \S^{-1/2} \V^* = \U^* \D^*,
$$
for which $\Exp(\Zpcr) = \0$ and $\Var(\Zpcr) = {\D^*}^2/n = \diag({\lambda_i^*}^2/n)$, so the features are centred, uncorrelated and with decreasing variances equal to the eigenvalues of the marginal sample correlation matrix, $\S^{-1/2}\hat{\Sigmab}\S^{-1/2}$. 

\subsection{Scaled SVD projection}

The scaled SVD transformation (\code{SVD}) is computed as:
$$
\Zsvd \;\leftarrow\; \XX \S^{-1/2} \V^* {\D^*}^{-1/2} = \U^* {\D^*}^{1/2},
$$
for which $\Exp(\Zsvd) = \0$ and $\Var(\Zsvd) = \D^*/n = \diag(\lambda_i^*/n)$.
Again the features are centred, uncorrelated and with decreasing variances equal to the square root of the eigenvalues of the marginal sample correlation matrix, $\S^{-1/2}\hat{\Sigmab}\S^{-1/2}$. 
In this case the features' dispersion presents a gentle decline compared to the \code{PCR} case.

\subsection{Remarks}

All the above transformations, except for STD, allow one to remove the ``ordering'' effect, so that even in the presence of ties the partitions obtained by applying MBHAC are invariant to permutations of the input variables. 
Furthermore, as shown in the next section through examples with real data, most of them allow one to achieve better clustering results when used for initialising the EM algorithm in Gaussian model-based clustering. 

\begin{figure}[htb]
\centering
\includegraphics[width=\textwidth]{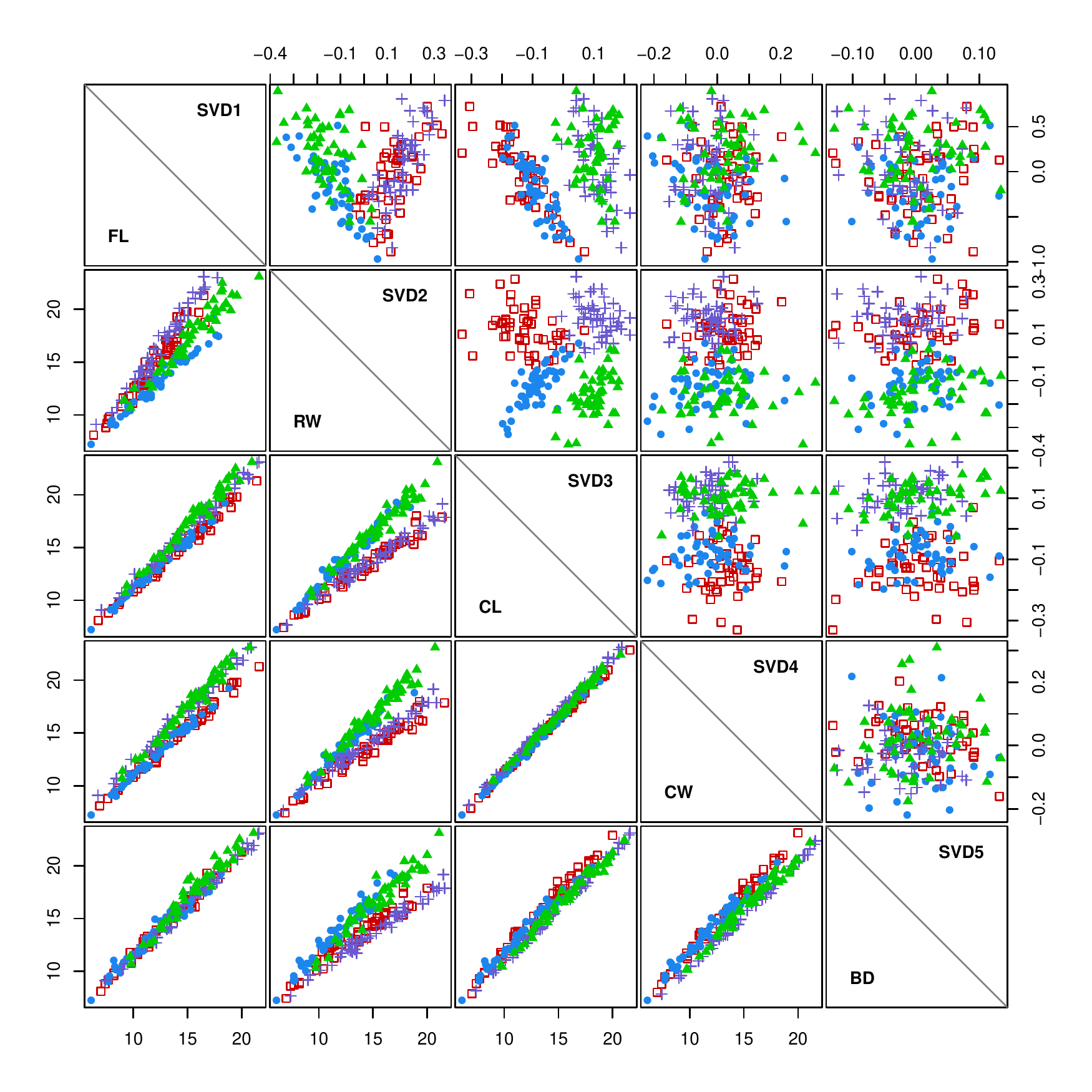}
\caption{Scatterplot matrix for the Crabs data: lower panels show scatterplots for pairs of variables in the original scale; upper panels show the features obtained by applying the scaled SVD transformation. For all graphs points are marked according to the true classes.}
\label{fig2:crabs}
\end{figure}

To get an idea of this, consider Figure~\ref{fig2:crabs} where we report a scatterplot matrix with pairs of plots for the original variables in the Crabs dataset (to be discussed in Section~\ref{sec:analyses}.1) and for scaled SVD-transformed features in, respectively, the lower and upper panels, and with points marked according to the true classes. In the graphs for the variables in the original scale is hard to detect any clustering structure, whereas in the first three SVD features there appears to be a certain degree of separation among classes. Applying MBHAC on this features space should provide better starting points for applying the EM algorithm.

\section{Data analyses}
\label{sec:analyses}

In this Section we present some examples using real data. We compare the behaviour of the proposed transformations against the usual MBHAC for starting the EM algorithm, and two common strategies for GMMs initialisation. The first strategy is to start from the best partition obtained out of 50 runs of the $k$-means algorithm. The second is the \textit{em}EM strategy as implemented in the \code{mixmod} software \citep{Biernacki:Celeux:Govaert:Langrognet:2006, Rpkg:Rmixmod}, which, by default, uses 50 short runs of EM, each made of 5 iterations, followed by a long run of EM from the solution maximising the log-likelihood.
The tolerance used for assessing the log-likelihood relative convergence of the EM algorithm is set to $10^{-5}$.

The comparison among the different initialising strategies is based on two measures, the BIC and the adjusted Rand index (ARI). 
The former is used to select the best Gaussian finite mixture model with respect to both the number of components and the component-covariances decomposition.
The BIC for model $\Model$ with $k$ components has the following form
$$
\BIC_{\Model,k} = 2\ell(\widehat{\PSI}; \x) - \nu_{\Model,k} \log(n),
$$
where $\ell(\widehat{\PSI}; \x)$ is the maximised log-likelihood, $\nu_{\Model,k}$ is the number of independent parameters to be estimated in model $\Model$,
 and $k$ is the number of mixture components. This criterion depends on the starting partition through the log-likelihood at the MLEs $\widehat{\PSI}$, penalised by the complexity of the model. It is the default criterion used in \pkg{mclust} for selecting a model, so the larger the value of the BIC the stronger the evidence for the corresponding model and number of components.

The ARI \citep{Hubert:Arabie:1985} is used for evaluating the clustering obtained with a given mixture model. This is a measure of agreement between two partitions, one estimated by a statistical procedure independent of the labelling of the groups, and one being the true classification. The ARI has zero expected value in the case of a random partition, and it is bounded above by 1, with higher values representing better partition accuracy.
Furthermore, it can be applied to compare partitions having different numbers of parts. 
The ARI is the index recommended by \citet{Milligan:Cooper:1986} for measuring the agreement between an external reference partition and a clustering partition. In the following data analysis examples, we take advantage of the knowledge of the true classes for measuring clustering accuracy, but not for model fitting. 

\subsection{Crabs data}
\begin{figure}[htb]
\centering
\includegraphics[width=0.7\textwidth]{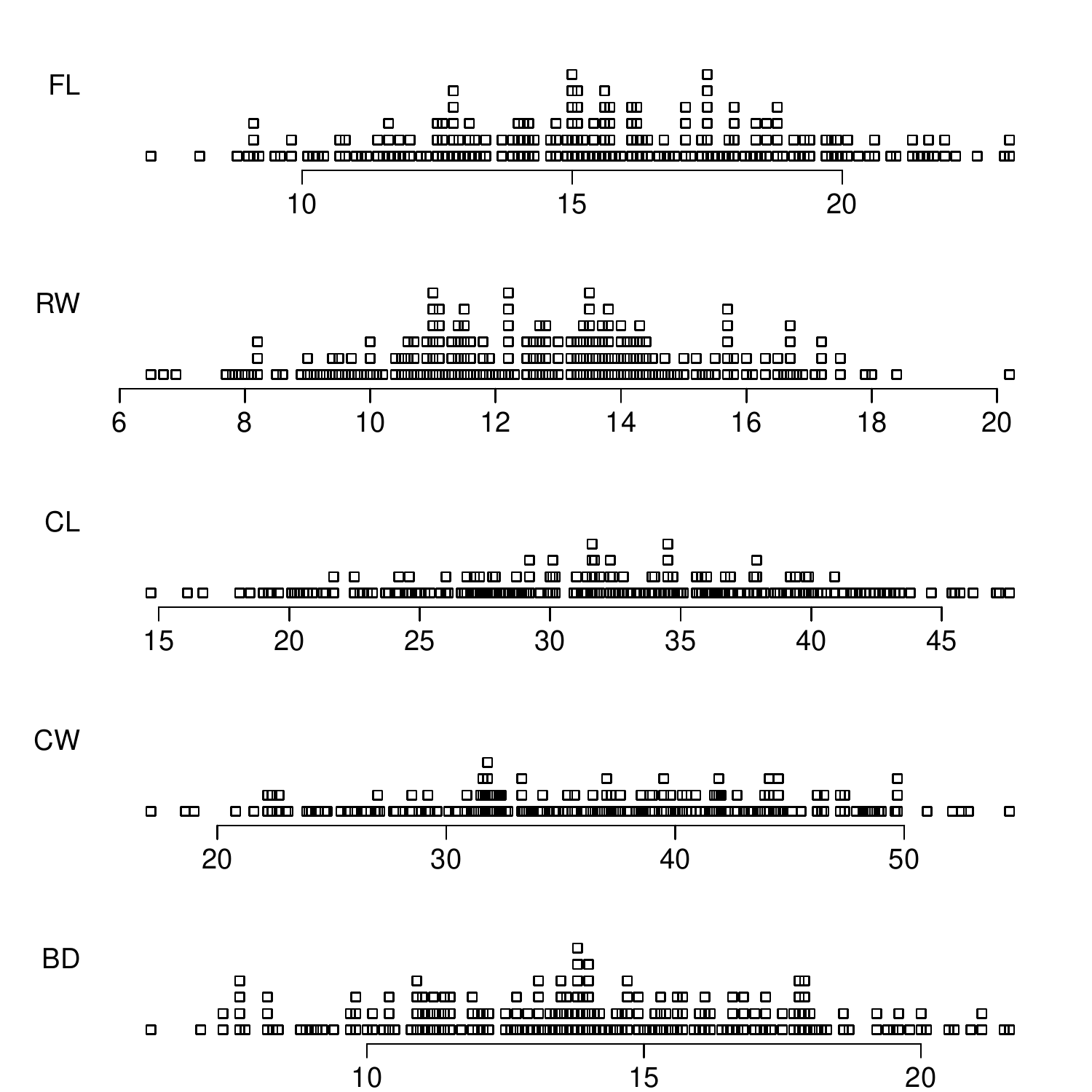}
\caption{Stripcharts with stacked data points for the Crabs data showing the marginal distribution for each variable. From this plot the presence of several ties is clearly visible.}
\label{fig1:crabs}
\end{figure}

We now consider a dataset consisting of data on five morphological measurements for 200 Leptograpsus crabs, with 50 crabs for each of two colour forms (blue and orange) and both sexes. Figure~\ref{fig1:crabs} shows the marginal distribution for each variable. In each stripchart data points are stacked to avoid overalapping, so the presence of several ties in the data is evident.

Overall, there are $5!=120$ possible different ordering of the variables, of which 105 lead to selection of the (EEE,9) model, and 15 the (EEV,3) model. Table~\ref{tab1:crabs} shows the corresponding BIC and ARI obtained for these models and for models initialised using the transformations discussed in Section~\ref{sec:mbhact}. As it can be seen, initialisation with the scaled SVD transformation returns the model with both the largest BIC and the most accurate partition (ARI = 0.7938). This transformation is the only one that selects the correct number of mixture components. 
Initialisation based on $k$-means also yields the right number of components, but its fit and accuracy are substantially worse. The \textit{em}EM initialisation strategy yields a better fit, but it selects the wrong number of clusters.

\begin{table}[htb]
\caption{Results from different initialisation strategies for the Crabs data. All possible orderings of the variables are reported for the default MBHAC initialisation; values among parentheses provide the number of such orderings which converge to the same solution. System times are averaged over 10 runs of the estimation algorithm.}
\label{tab1:crabs}
\centering
\begin{tabular}{lccrrr}
\toprule
Initialisation  & $\Model$ & $k$ & BIC & ARI & System time \\
\midrule
MBHAC & & & & & \\
\cline{1-1} 
Default (105)          & EEE & 9 & -2883.68 & 0.4831 & 2.61 s \\
\phantom{Default} (15) & EEV & 3 & -2925.59 & 0.4715 & 2.47 s \\
SPH             & EEE &  8 & -2861.27 & 0.5248 & 2.40 s \\
PCS             & EEE &  9 & -2883.68 & 0.4831 & 2.69 s \\
PCR             & EEE & 10 & -2894.65 & 0.4572 & 2.72 \\
SVD             & EEV &  4 & \textbf{-2842.30} & \textbf{0.7938} & 1.88 s \\
\midrule
$k$-means       & EEV & 4 & -2916.68 & 0.5926 & 2.78 s \\
\textit{em}EM   & EEE & 6 & -2866.61 & 0.6305 & 11.08 s \\
\bottomrule
\end{tabular}
\end{table}

\citet{Raftery:Dean:2006} selected as optimal subset for clustering purposes the variables \code{(FL, RW, CW, BD)} (in the identified ordering). Also for such a subset, different orderings give different results in 7 out of 24 possible arrangements.
Table~\ref{tab2:crabs} shows the clustering results obtained with different initialisation strategies using the above mentioned optimal subset of variables. Again, initialisation with the scaled SVD transformation yields the largest BIC and the highest ARI among the considered strategies, with a slight improvement over the default initialisation with the identified ordering of the variables. Note that \textit{em}EM has results analogous to the best default MBHAC initialisation, but $k$-means results are markedly worse.

\begin{table}[htb]
\caption{Results from different initialisation strategies for the Crabs data using the optimal subset of variables identified by \citet{Raftery:Dean:2006}. All possible orderings of the variables are reported for the default MBHAC initialisation; values among parentheses provide the number of such orderings which converge to the same solution. System times are averaged over 10 runs of the estimation algorithm.}
\label{tab2:crabs}
\centering
\begin{tabular}{lccrrr}
\toprule
Initialisation          & $\Model$ & $k$  & BIC & ARI & System time \\
\midrule
MBHAC & & & & & \\
\cline{1-1} 
Default (17)          & EEV &  4 & -2609.89 & 0.8154 & 2.54 s\\
\phantom{Default} (5) & EEV &  5 & -2645.19 & 0.7487 & 2.40 s\\
\phantom{Default} (1) & EEV &  4 & -2609.90 & 0.8276 & 2.40 s\\
\phantom{Default} (1) & EEE & 11 & -2674.40 & 0.4106 & 2.46 s\\
SPH          & EEV &  4 & -2609.79 & 0.8154 & 1.89 s\\
PCS          & EEV &  4 & -2609.89 & 0.8154 & 2.55 s\\
PCR          & EEE &  8 & -2652.58 & 0.4860 & 2.52 s\\
SVD          & EEV &  4 & \textbf{-2609.78} & \textbf{0.8400} & 2.06 s\\
\midrule
$k$-means     & EEE &  5 & -2661.07 & 0.5608 & 2.62 s\\
\textit{em}EM & EEV &  4 & -2609.87 & 0.8154 & 8.70 s\\
\bottomrule
\end{tabular}
\end{table}


Finally, note that, using both the full set of the variables and the optimal subset, analysis with the initialisation using the scaled SVD transformation used the smallest computing time. This may appear counterintuitive at first sight, because computing the SVD is time consuming. However, this is more than balanced by the fact that having good starting values allows the EM algorithm to converge in fewer steps.

\subsection{Flea beetles data}

This dataset provides a sample of 72 flea beetles from three species. We consider the clustering using the six physical measurements available, namely \code{(tars1, tars2, head, aede1, aede2, aede3)}.

There are $6!=720$ different orderings of the variables.  
A huge amount of variability is obtained by using the default initialisation: about half the times the (EEE,3) model is selected, but also models with 4, 5, and 6 components are chosen in a non negligible number of cases. Even when the same pair of covariance matrix decomposition and number of components is selected, the estimated models are different. In fact, there are 104 unique BIC values (rounding to second decimal point). A picture of these results is shown in the Figure appearing in Table~\ref{tab1:flea}, with the area of dots proportional to the observed counts. 
Table~\ref{tab1:flea} contains the results for some orderings of the variables, and for the transformation-based initialisations discussed in Section~\ref{sec:mbhact}. 
From these we can see there are several reasonable starting values that lead to the best fitting model, which is also able to recover the true partition. Using any transformation-based initialisation, except that based on PCS, allows to recover the true partition without requiring more computing time, and removing the uncertainty connected with the ordering of variables.
 
\begin{table}[htb]
\centering
\caption{Results from different initialisation strategies for the Flea beetles data. Default initialisation reports the results obtained with some orderings of the variables; values among parentheses provide the number of such orderings which converge to the same pair $(\Model,k)$. System times are averaged over 10 runs of the estimation algorithm.}
\label{tab1:flea}
\setlength\tabcolsep{0.6ex}
\begin{tabular}{lccrrr}
\toprule
Initialisation & $\Model$ & $K$ & BIC & ARI & System time \\
\midrule
MBHAC & & & & & \\
\cline{1-1} 
Default (367) & EEE & 3 & \textbf{-2785.57} & \textbf{1.0000} & 0.17 s \\
\phantom{Default} (209) & EEE & 4 & -2806.80 & 0.9664 & 0.21 s \\
\phantom{Default} (123) & EEE & 5 & -2810.68 & 0.7886 & 0.22 s \\
\phantom{Default}  (20) & EEE & 6 & -2838.47 & 0.5642 & 0.19 s  \\
\midrule
STD          & EEE & 3 & \textbf{-2785.57} & \textbf{1.0000} & 0.13 s \\
SPH          & EEE & 3 & \textbf{-2785.57} & \textbf{1.0000} & 0.15 s \\
PCS          & EEE & 4 & -2802.75 & 0.7886 & 0.17 s \\
PCR          & EEE & 3 & \textbf{-2785.57} & \textbf{1.0000} & 0.12 s \\
SVD          & EEE & 3 & \textbf{-2785.57} & \textbf{1.0000} & 0.17 s \\
\midrule
$k$-means     & EEE & 3 & \textbf{-2785.57} & \textbf{1.0000} & 0.17 s \\
\textit{em}EM & EEE & 3 & \textbf{-2785.57} & \textbf{1.0000} & 0.17 s \\
\bottomrule
\end{tabular}
\end{table}

\subsection{Female voles data}

\cite{Flury:1997} reported the data for a sample of 86 female voles from two species, 41 from Microtus californicus and 45 from Microtus ochrogaster, and seven variables describing various body measurements in units of 0.1mm, so several ties are contained in this data set.

There are $7!=5{,}040$ possible orderings of the variables. 
When the default initialisation is used, there is considerable variation
among the resulting solutions:
41 final models with different BIC values (up to 5 significant digits) are estimated, leading to 38 different partitions. About 67\% of the solutions give a single component model, about 20\% have three components, and only about 7.5\% of the solutions correctly identify the correct number of clusters.
On the contrary, three transformation-based initialisation strategies are able to achieve the best fit in term of BIC, which in turn provides the largest ARI. The same optimal solution is also achieved by the \textit{em}EM strategy, whereas results from the $k$-means initialisation are inferior.
Note that also in this case the scaled SVD transformation is among the best strategies, with no increase in computing time.

\begin{table}[htb]
\centering
\caption{Results from different initialisation strategies for the Female voles data. The results for the default MBHAC initialisation are reported for some of all possible orderings of the variables; values among parentheses provide the number of such orderings which converge to the same pair $(\Model, k)$. 
System times are averaged over 10 runs of the estimation algorithm.}
\label{tab1:fvoles}
\setlength\tabcolsep{2ex}
\begin{tabular}{lccrrr}
\toprule
Initialisation & $\Model$ & $K$ & BIC & ARI & System time \\
\midrule
MBHAC & & & & & \\
\cline{1-1} 
Default (3395) & XXX & 1 & -3874.9 & 0.0000 & 0.19 s \\
\phantom{Default} (996) & EEE & 3 & -3862.8 & 0.7720 & 0.29 s \\
\phantom{Default} (381) & EEE & 2 & \textbf{-3844.2} & \textbf{0.9081} & 0.23 s \\
\phantom{Default} (253)        & EEE & 4 & -3872.6 & 0.6015 & 0.24 s \\
SPH          & EEE & 5 & -3871.0 & 0.4129 & 0.26 s \\
PCS          & XXX & 1 & -3874.9 & 0.0000 & 0.22 s \\
PCR          & EEE & 2 & \textbf{-3844.2} & \textbf{0.9081} & 0.22 s \\
SVD          & EEE & 2 & \textbf{-3844.2} & \textbf{0.9081} & 0.19 s \\
\midrule
$k$-means     & EEE & 4 & -3872.6 & 0.6015 & 0.25 s \\
\textit{em}EM & EEE & 2 & \textbf{-3844.2} & \textbf{0.9081} & 2.43 s \\
\bottomrule
\end{tabular}
\end{table}


\subsection{Italian wines data}

\citet{Forina:1986} reported data on several chemical and physical properties of 178 wines grown in the same region in Italy but derived from three different cultivars (Barolo, Grignolino, Barbera). 
We consider 27 of the original 28 variables that were described in the Forina et. al paper and are available in the \pkg{pgmm} \proglang{R} package \citep{Rpkg:pgmm}. In what follows, the data are analysed on the standardised scale and, given the large number of features, only the common full covariance matrix model (EEE) is examined.
Table~\ref{tab1:wines} shows the results obtained using different MBHAC initial partitions, $k$-means and \textit{em}EM initialisation strategies. Except for one case, the MBHAC initialisations allow one to achieve good clustering accuracy. 
In particular, SVD has the highest BIC and attains a clustering that perfectly matches the real classes. By comparison, the models initialised by $k$-means and \textit{em}EM show somewhat worse values of both BIC and ARI.

\begin{table}[htb]
\centering
\caption{Results from different initialisation strategies for the Italian wines data. System times are averaged over 10 runs of the estimation algorithm.}
\label{tab1:wines}
\setlength\tabcolsep{2ex}
\begin{tabular}{lccrrr}
\toprule
Initialisation & $\Model$ & $K$ & BIC & ARI & System time \\
\midrule
MBHAC & & & & & \\
\cline{1-1}
Default & EEE & 3 & -12309.90 & 0.9489 & 0.36 s \\
SPH  & EEE & 2 & -12370.80 & 0.5820 & 0.78 s \\
PCS  & EEE & 3 & -12309.90 & 0.9489 & 0.36 s \\
PCR  & EEE & 3 & -12309.90 & 0.9489 & 0.37 s \\
SVD  & EEE & 3 & \textbf{-12306.75} & \textbf{1.0000} & 0.35 s \\ 
\midrule
$k$-means     & EEE & 3 & -12314.72 & 0.9637 & 0.74 s \\
\textit{em}EM & EEE & 4 & -12312.04 & 0.8355 & 3.99 s \\
\bottomrule
\end{tabular}
\end{table}

\section{Final comments}
\label{sec:final}

The mixture model-based approach to clustering provides a firm statistical framework for unsupervised learning. The EM algorithm is typically used for parameter estimation. However, because of the many local maxima of the likelihood function, an important problem for getting sensible maximum likelihood estimates is to obtain reasonable starting points. A computationally efficient approach to EM initialisation is based on the partitions obtained from agglomerative hierarchical clustering. 

In this paper we have presented and experimented with simple transformation-based methods to refine the EM initialisation step derived from model-based agglomerative hierarchical clustering. The proposed transformation-based strategies allow one to remove the dependence on the ordering of the variables when selecting starting partitions. They also often lead to improved model fitting and more accurate clustering results. Among the investigated transformations, the scaled SVD transformation performed the best in our experiments.
The proposed approach may be applicable to other mixture modelling contexts, but we have not explored this possibility yet. Future studies will be devoted to this aspect.

As mentioned by \citet[p. 567]{Biernacki:etal:2003} and \citet{Melnykov:Maitra:2010}, we cannot expect an initialisation strategy to work uniformly well in all cases. Therefore, it is important to explore different strategies and to choose the solution with the highest log-likelihood value, but also to consider those sub-optimal solutions on a more subject specific considerations.

The transformation-based initialisation strategies discussed in this paper are available in the \proglang{R} package \pkg{mclust} (version $>= 4.4$), which can be downloaded from CRAN at \url{http://cran.r-project.org/web/packages/mclust/index.html}.

\bibliographystyle{spbasic}
\bibliography{mbhact}

\end{document}